\documentclass[aos,preprint]{imsart}

\usepackage[hang]{subfigure}
\usepackage{color}
\usepackage{latexsym}
\usepackage{mathtools}
\usepackage{amsmath}
\usepackage{amssymb}
\usepackage[utf8]{inputenc}
\usepackage{textcomp}
\usepackage{graphicx}
\usepackage{epsfig}
\usepackage{dsfont}
\usepackage{hyperref}
\usepackage{bbm}
\usepackage{enumitem}
\usepackage{here}

\usepackage{color}
\usepackage[dvipsnames]{xcolor}

\RequirePackage{amsthm,amsmath,amsfonts,amssymb}
\RequirePackage[authoryear,numbers]{natbib}
\RequirePackage[colorlinks,citecolor=blue,urlcolor=blue]{hyperref}
\RequirePackage{graphicx}

\startlocaldefs

\renewcommand{\epsilon}{\varepsilon}

\newcommand{\Z}{{\mathbb Z}}

\newcommand{\R}{\mathbb R}

\newcommand{\bea}{\begin{eqnarray*}}
\newcommand{\eea}{\end{eqnarray*}}
\newcommand{\be}{\begin{eqnarray}}
\newcommand{\ee}{\end{eqnarray}}
\newcommand{\ba}{\begin{array}}
\newcommand{\ea}{\end{array}}

\newcommand{\beq}{\begin{equation}}
\newcommand{\eeq}{\end{equation}}
\newcommand{\bi}{\begin{itemize}}
\newcommand{\ei}{\end{itemize}}
\newcommand{\bpm}{\begin{pmatrix}}
\newcommand{\epm}{\end{pmatrix}}

\newcommand{\proc}{\{X_t :  t\in \Z\}}

\newcommand{\spec}{\boldsymbol{\mathfrak{f}}}
\newcommand{\specdis}{\boldsymbol{\mathfrak{F}}}

\newcommand{\copper}{\mathcal{I}}

\theoremstyle{plain}

\theoremstyle{remark}

\newcommand{\dd}{\mathrm{d}}

\usepackage{tikz}
\usepackage{pgfplots}
\newcommand{\dashedtriangle}{\raisebox{0pt}{\tikz{\draw[red,solid,line width=1pt] (0,0) -- (1mm,1.7mm) -- (2mm,0) -- cycle; \draw[red,dashed,line width=1pt] (-2.5mm,0.85mm) -- (4.5mm,0.85mm);}}}

\newcommand{\solidcircle}{\raisebox{0pt}{\tikz{\draw[black,solid,line width=1pt] (0,0) circle (0.8mm); \draw[black,solid,line width=1pt] (-3mm,0) -- (3mm,0);}}}

\newcommand{\dottedplus}{\raisebox{0pt}{\tikz{\draw[green,solid,line width=1pt] (-0.8mm,0) -- (0.8mm,0); \draw[green,solid,line width=1pt] (0,-0.8mm) -- (0,0.8mm);
\draw[green,dotted,line width=1pt] (-3mm,0) -- (3mm,0);}}}

\endlocaldefs

\begin{document}

\begin{frontmatter}
	\title{A model-free test of the time-reversibility \\ of climate change processes}
	\runtitle{Time-irreversibility of climate change processes}
	
	\begin{aug} 
\author[A]{\fnms{Yuichi} \snm{Goto}\ead[label=e1]{yuichi.goto@math.kyushu-u.ac.jp}},
\author[F,B]{\fnms{Marc}
\snm{Hallin}\ead[label=e6]{mhallin@ulb.be}}

\address[A]{\qq{Faculty of Mathematics}, Kyushu University\printead[presep={,\ }]{e1}}
\address[F]{\qq{ECARES} and \qq{D\' epartement de Math\' ematique}, Universit\' e libre de Bruxelles\printead[presep={,\ }]{e6}}
\address[B]{\qq{Institute of Information Theory and Automation}, Czech Academy of Sciences, Prague, Czech Republic}\vspace{1mm}


\end{aug}

\begin{abstract}
Time-reversibility is a crucial feature of many   time series models, while time-irreversibility is the rule rather than the exception in real-life data. Testing the null hypothesis of time-reversibilty, therefore, should be an important step preliminary to the identification and estimation of most traditional time-series models.  Existing procedures,  however, mostly consist of testing necessary but not sufficient conditions, leading to under-rejection, or sufficient but non-necessary ones, which leads to over-rejection. Moreover, they generally are model-besed. In contrast, the copula spectrum  studied by Goto et al.~(\textit{Ann. Statist.} 2022,~\textbf{50}: 3563--3591) allows for a model-free necessary and sufficient time-reversibility condition. A test based on this copula-spectrum-based characterization  has been proposed by authors. This paper illustrates the performance of this test, with an  illustration in the analysis of  climatic data. 
\end{abstract}
\end{frontmatter}

\section{Introduction}

Time-reversibility (sometimes referred to as {\it lag-reversibility}  or {\it non-directionality}: see, e.g., \cite{Lawrance91}) is a crucial feature of  most classical  time-series models---a feature which, however,  is challenged by many real-life processes. Examples include observations from various fields: unemployment rates and the business cycle  \citep{neftci84}, electroencephalographic and heartbeat recordings \citep{VderH96, Palus96, Costaetal05},  epidemic outbreaks \citep{Grenfelletal94}, ... to quote only a very few.

This observation has prompted researchers to investigate time-reversibility more closely.
 \cite{weiss75} showed that 
 (i) a stationary Gaussian process is always time-reversible; 
 (ii) for stationary ARMA($p$,$q$) with $p\geq1$, time-reversibility and Gaussianity are equivalent;  
(iii)  the same equivalence holds for MA($q$) processes with non-symmetric  or non-antisymmetric coefficients  (a MA($q$) process of the form $X_t=\sum_{i=0}^q b_i\varepsilon_{t-i}$ with i.i.d.~$\varepsilon_t$'s has symmetric or anti\-symmetric coefficients if $b_i=\pm b_{q-i}$, $i=0,\ldots,q$);
(iv) MA($q$) processes with symmetric coefficient ($b_i=  b_{q-i}$) are always time-reversible (irrespective of the  density of the i.i.d.~$\varepsilon_t$'s);
(v) for MA($q$) processes with antisymmetric coefficients ($b_i=- b_{q-i}$), time-reversibility holds if the  density of the i.i.d.~$\varepsilon_t$'s is symmetric.
\cite{hcp88}, \cite{bd92}, \cite{cheng99} extended   Weiss' results to MA($\infty$) processes. 
\cite{tc2005} and \cite{cht2006} studied time-reversibility for vector linear processes. 
\cite{McK85} and \cite{lmh89} introduced, under the terminology 
{\it positively} or {\it negatively correlated pairs of beta random variables} (PBAR and NBAR, respectively), classes of  auto\-regressive beta-gamma processes which are non-linear but time-reversible. 

Several approaches have been developed to test the null hypothesis of time-reversibility.
\cite{rr96} propose (under moment assumptions) a test based on the fact that, for a time-reversible process $\{X_t\}$, ${\rm E}X_t^iX_{t-k}^j={\rm E}X_t^jX_{t-k}^i$ for all $i,j,k\in\mathbb N$. \cite{giannakis1994} introduce a test for time-reversibility based on  third-order cumulants. \cite{br67} point out that, for time-reversible process, the imaginary part of any higher-order spectrum vanishes. The converse is true, provided that the distribution of the process is determined by its moments. Tests based on bi- and tri-spectra are considered by \cite{hr98} and \cite{wfh2014}.
\cite{cck2000}, \cite{ck2002}, \cite{pp2002}, \cite{chen2003}, \cite{rm2007}, while \cite{psaradakis2008} proposes  tests based on the fact that, for any time-reversible~$\proc$, the  process~
$\{Y_{t,k}\coloneqq X_t-X_{t-k}: t\in\Z\}$ is symmetric about the origin for all $k\in\mathbb N$ . 
\cite{proietti2020} constructs a test based on the transition probability of the process associated with a maximum process. 
\cite{dvtd1995} and \cite{bs2014} propose a test based on  joint distribution functions and 
\cite{dfg2004} a test based on  canonical directions. 
\cite{smp2009} propose a test for time-reversibility of stationary finite-state Markov chains. We refer to \cite{Kathpaliaetal21} or 
Chapter 8 of \cite{goojier2017} for  a comprehensive review of the problem.

Many of these tests are based on necessary but not sufficient conditions for time-reversibility, which unavoidably  induces  blind spots, hence biasedness and under-rejection. Sufficient-condition-based tests similarly induce over-rejection. Tests based on necessary and sufficient conditions, thus, are highly desirable. 
This is what the copula spectrum 
 develo\-ped by \cite{gkvvdh22}, which offers a model-free  characterization of pairwise time-reversibility, is allowing for.  
In this paper, we illustrate the performance of the copula-spectrum-based tests proposed by \cite{gkvvdh22}
 in comparison with the recent model-based approach by \cite{ghm22}, who are using it  to assess the time-reversibility of climate-related time-series data.

The  paper is organized as follows: Section \ref{sec::TR} reviews the approaches developed by \cite{gkvvdh22} and by \cite{ghm22}, respectively. Section \ref{sec::NS} presents our numerical results, and Section \ref{sec::ES} compares the  conclusions of the two approaches on the climate-related application considered in   \cite{ghm22}.

\section{Testing for time-reversibility}\label{sec::TR}
A strictly stationary process $\proc$ is said to be {\it pairwise time-reversible} if the distribution of $(X_t,X_{t+k})$ is equal to that of $(X_{t+k},X_t)$ for all~$k\in\mathbb Z$.
We are interested in the pairwise time-reversibility of the process and consider testing  
\[H_0: \proc\text{ is pairwise time-reversible}\]
against
\[H_1:\ \proc\text{ is not pairwise time-reversible}
.\]
In the sequel, for the sake of simplicity, we write  {\it time-reversibility} and  {\it time-reversible} for   {\it pairwise time-reversibility} and  {\it pairwise time-reversible}. A process $\proc$ which  is not pairwise time-reversible will be referred to as {\it pairwise time-irreversible}---in short,  {\it time-irreversible} or {\it non-reversible}.

\subsection{A test based on the  integrated copula spectrum}
A test for time-reversibility based on the  {\it integrated copula spectrum} (hereafter referred to as~ICS) with critical values obtained via a subsampling method was proposed by \cite{gkvvdh22}.

The ICS is defined, for a strictly stationary process $\proc$ with  marginal distribution function $F$, as
\begin{align*}
\specdis(\lambda ;\tau_1,\tau_2)\coloneqq \int_0^{\lambda}\spec(\omega; \tau_1,\tau_2)\dd\omega , \qquad(\tau_1,\tau_2)\in (0,1)^2,\ \ \lambda\in [0,\pi],
\end{align*}
where $\mathfrak{f}$ is the {\it copula spectral density},  given by ($I\{A\}$ stands for the indicator  of $A$)
\begin{align*}
\mathfrak{f}(\omega; \tau_1,\tau_2)\coloneqq \frac{1}{2\pi}\sum_{k \in \Z}{\text{\rm Cov}}(I\{F(X_k) \leq \tau_1\}, I\{F(X_0) \leq \tau_2\})e^{-i\omega k},\quad (\omega,\tau_1,\tau_2)\in \R\times(0,1)^2\!.
\end{align*}
 As an estimator of $\specdis$, we propose 
\begin{align*}
	\widehat{\specdis}\phantom{F\!\!\!\!\!}_{n,R}(\lambda ;\tau_1,\tau_2)\coloneqq {}&\frac{2\pi}{n}\sum_{s=1}^{n-1}I\big\{0\leq \frac{2\pi s}{n}\leq\lambda\big\}\copper_{n,R}^{\tau_1,\tau_2}\big(\frac{2\pi s}{n}\big),\qquad\lambda\in [0,\pi],
\end{align*}
where 
\begin{align*}
	\copper_{n,R}^{\tau_1,\tau_2}(\omega)\coloneqq \frac{1}{2\pi n}d_{n,R}^{\tau_1}(\omega)d_{n,R}^{\tau_2}(-\omega),\qquad (\omega,\tau_1,\tau_2)\in \R\times(0,1)^2
\end{align*}
with
\begin{align*}
	d_{n,R}^{\tau}(\omega)\coloneqq \sum_{t=0}^{n-1}I\{\hat{F}_n(X_t)\leq\tau\}e^{-i\omega t}\text{ and }
	\hat{F}_n(x)\coloneqq \frac{1}{n}\sum_{t=0}^{n-1}I\{X_t\leq x\}
\end{align*}
  is the {\it copula rank periodogram}. \cite{gkvvdh22} show that $\proc$ is   pairwise time-reversible if and only if 
{the imaginary part $\Im\specdis(\lambda ;\tau_1,\tau_2)$ of $\specdis(\lambda ;\tau_1,\tau_2)$ vanishes for all~$(\lambda,\tau_1, \tau_2)\in[0,\pi]\times(0,1)^2$}. 

Our test statistic accordingly is defined as
\begin{align*}
T_{\rm TR}^{(n)}\coloneqq &\sqrt{n} \max_{(\lambda, \tau_1, \tau_2) \in \mathcal T}\Big| {\Im \widehat{\specdis}\phantom{F\!\!\!\!\!}_{n,R}(\lambda, \tau_1, \tau_2)}\Big|,
\end{align*}
where $\Im \widehat{\specdis}\phantom{F\!\!\!\!\!}_{n,R}$ denotes the imaginary part of $\widehat{\specdis}\phantom{F\!\!\!\!\!}_{n,R}$ and 
\[\mathcal T\coloneqq \{2\pi \ell/32; \ell = 0,1,\ldots,16\}\times\{j/32; j=1,\ldots,31\}^2\] 
is a discrete grid {(16337 gridpoints)} over~$[0,\pi]\times(0,1)^2$. The p-value $p_{{\rm TR}}$ is calculated via the subsampling procedure:
\begin{equation}\label{p-val}
p_{{\rm TR}}\coloneqq  \frac{1}{n-b+1} \sum_{t=0}^{n-b}I\big\{T_{{\rm TR}}^{(n,b,t)}> T_{\rm TR}^{(n)}\big\}, 
\end{equation}
where
\begin{align*}
&T_{\rm TR}^{(n,b,t)}\coloneqq  \left(1 - \frac{b}{n}\right)^{-1/2}\sqrt{b}\max_{ (\lambda, \tau_1, \tau_2) \in \mathcal T}\Big| {\Im \widehat{\specdis}\phantom{F\!\!\!\!\!}_{n,b,t,R}(\lambda, \tau_1, \tau_2)}\Big|,\\
&\widehat{\specdis}\phantom{F\!\!\!\!\!}_{n,b,t,R}(\lambda ;\tau_1,\tau_2)
\coloneqq \frac{2\pi}{b} \sum_{j=1}^{b-1} I\big\{0\leq \frac{2\pi j}{b}\leq\lambda\big\} \copper_{n,b,t,R}^{\tau_1,\tau_2}\Big( \frac{2\pi j}{b} \Big),\\
&	\copper_{n,b,t,R}^{\tau_1,\tau_2}(\omega)\coloneqq \frac{1}{2\pi b}d_{n,b,t,R}^{\tau_1}(\omega)d_{n,b,t,R}^{\tau_2}(-\omega),\qquad (\omega,\tau_1,\tau_2)\in \R\times(0,1)^2,\\
&	d_{n,b,t,R}^{\tau}(\omega)\coloneqq \sum_{j=0}^{b-1}I\{\hat{F}_{n,b,t}(X_{t+j})\leq\tau\}e^{-i\omega j}, 
	\end{align*}
and
$$\widehat F_{n,b,t}(x) \coloneqq  \frac{1}{b} \sum_{i=t}^{t+b-1} I\{X_{i} \leq x\};
$$
the block length $b$ is determined by the rule of thumb
\begin{equation}\label{eqn:rt_bw}
b^{\rm rt}_n \coloneqq  \max\{ 2^j :  2^j \leq 2n^{2/3}, \ j=4,\ldots,8 \}.
\end{equation}
The test rejects $H_0$  at significance level  $\alpha$ whenever $p_{{\rm TR}}<\alpha$.
Under mild regularity conditions,  ICS-based tests have asymptotically correct size and are consistent against any fixed alternative in $H_1$: see Theorem 4.2 in  \cite{gkvvdh22}.
\medskip

\subsection{The \cite{ghm22} test}
\cite{ghm22} recently proposed a model-based testing procedure (hereafter referred to as GHM)  for the same problem. Their test involves the following steps.  
\begin{enumerate}
\item[] \textbf{Step 1}  Fit  a causal AR($p$) model to the data (order $p$ determined by BIC).
\item[] \textbf{Step 2} Apply the Shapiro-Wilk or/and the Jarque-Bera test \citep{ShapWilk65,JBera80} of normality to the residuals of the fitted AR($p$) model. If the hypothesis of normality   is not rejected, conclude that the data is time-reversible; otherwise, proceed to Step~3.
\item[] \textbf{Step 3} Considering an  AR($r+s$) model with unspecified mean $\mu$, $r$ causal roots, and~$s$ non-causal ones, driven by scaled i.i.d.~Student noise\footnote{The driving noise is of the form $\sigma\varepsilon_t$, where $\sigma >0$  is an unspecified scale and $\varepsilon_t$ is i.i.d.\ Student with unspecified degrees of freedom $\nu$.} with unspecified scale $\sigma>0$ and unspecified degrees of freedom $\nu$, estimate $\mu$, $\nu$,  $\sigma^2$, and the unknown AR parameters by Student  maximum  likelihood under the constraint   that $ r + s = p$ if $p$ is even, and $ r + s =p + 1$ if $p$ is odd, where $p$ is determined in Step 1. 
If the estimators of $r$ and $s$ differ,\footnote{Since $r$ and $s$ and their estimators $\hat r$ and $\hat s$ are integers, consistency implies that, for $n$ large enough, $\hat{r}=r$ and $\hat{s}=s$.} conclude that the data is time-irreversible; otherwise, proceed to Step~4. 
\item[] \textbf{Step 4} Based on a further constrained Student  likelihood with the additional constraint that~$r=s$   and the $r=s$ causal and non-causal AR parameters are equal,  re-estimate~$\mu$,~$\nu$,~$\sigma^2$, and the unknown AR parameters and proceed to one of the following steps. 
\item[] \textbf{Step 5 (Strategy 1)} 
If the BIC  for the unrestricted likelihood in Step~3 is strictly smaller than that for the restricted likelihood in Step~4, conclude that the series is time-irreversible; otherwise, conclude that  it is time-reversible.
\item[] \textbf{Step 5 (Strategy 2)} 
Based on the Student likelihoods computed in  Steps~3 and~4, perform a likelihood ratio test for the hypothesis that the causal and non-causal parameters are equal. If the hypothesis is rejected, conclude that the data is time-irreversible; otherwise, conclude that it is time-reversible.
\end{enumerate}

This procedure is it entirely heuristic---no consistency results are proven in  \cite{ghm22}---and its global size, when all steps are performed at nominal size $\alpha$, is anything but  clear; however, it seems to work well, at least when the data-generating process is linear.

\section{Numerical study}\label{sec::NS}

This section presents the respective simulation-based finite-sample performance of the ICS and   GHM methods, with a comparative discussion. 

In our simulations, we considered both positively and negatively correlated  Beta random variables autoregressive processes (PBAR and NBAR, see \cite{McK85}) of the form
\begin{align}\label{BAR}
X_t = 1-U_t(1-W_t X_{t-1})\quad\text{and}\quad
X_t = U_t(1-W_t X_{t-1}),\quad t\in\Z ,
\end{align}
where  $\{U_t\}$ and $\{W_t\}$, $t\in\Z$ are i.i.d.\ ${\rm Beta}(2,1)$ and ${\rm Beta}(1,1)$ random variables, respectively,  and $\{U_t\}$ is independent of $\{W_t\}$.  
The PBAR and NBAR models, proposed by \cite{McK85}, are non-linear and non-Gaussian but time-reversible, satisfying $H_0$. So, it is expected that the ICS and GHM procedures do not reject $H_0$. 

We also considered the quantile autoregressive (QAR) model proposed by \cite{Koenker2006}, which is an example of nonlinear, uncorrelated, and  time-irreversible   model, under which~$H_0$   should be rejected.  More precisely, in our simulations, we considered the~QAR(1) model 
\begin{align}\label{QAR}
X_t=0.1\Phi^{-1}(U_t)+1.9(U_t-0.5)X_{t-1},
\end{align}
where $\{U_t\}$ is  i.i.d.\   uniform over $[0,1]$ and $\Phi
$, as usual, stands for the   the standard normal distribution function. 

In the first step of the GHM procedure,   BIC is computed for $p=0,1,\ldots,5$. When applied to simulated QAR(1) series, BIC with high probability is  selecting $p=0$---that is, an i.i.d.\ AR(0) process---in which case  the GHM procedure logically should conclude to time-reversibility (although nothing is mentioned about this in \cite{ghm22}).  
%
The subsampling block length for the ICS procedure is determined by \eqref{eqn:rt_bw}. 

We applied the ICS  and   GHM (strategies 1 and~2) methods to data generated from the three models \eqref{BAR} and~\eqref{QAR} with series lengths~$n=100, 200, 500$, and~$1000$,   each with  1000 replications. All tests were performed at nominal level  $\alpha = 5\%$.

\begin{figure}[t]
	\begin{center}
\includegraphics[width = \linewidth, height = 0.34 \linewidth]{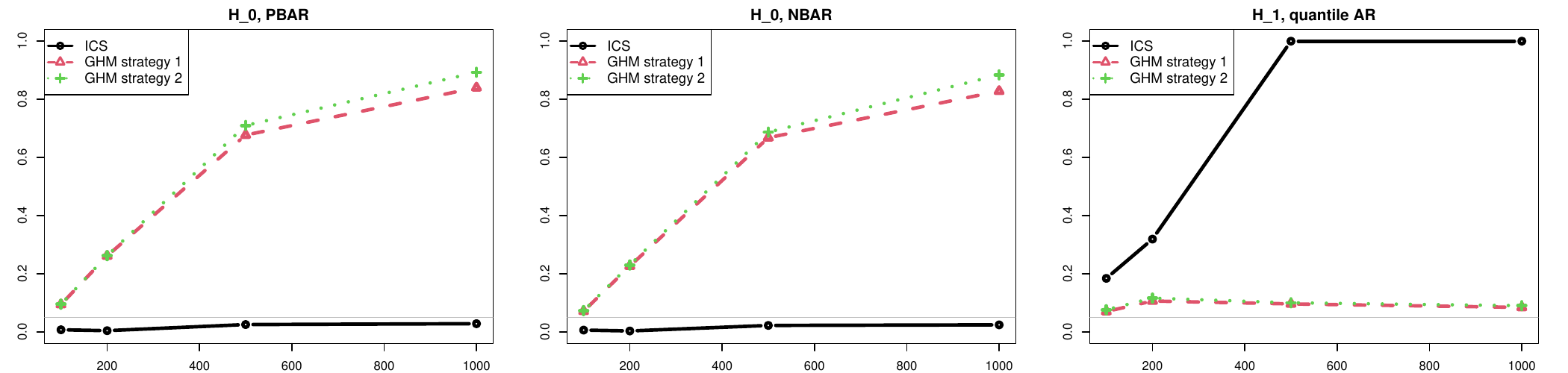}
	\end{center}
	\vspace{-0.4cm}
		\caption{\small 
Rejection frequencies (vertical axis) of the integrated copula-spectrum-based test (ICS, black solid line  \protect\solidcircle)  and the model-based tests (GHM  strategy 1, red dashed line \protect\dashedtriangle  and GHM  strategy~2, green  dotted line \protect\dottedplus) in 1000 replications of the PBAR (left panel; $H_0$ satisfied), NBAR (center panel; $H_0$ satisfied), and QAR (right panel; $H_0$ violated) processes with series lengths~$n=100, 200, 500$, and~$1000$ (horizontal axis). All tests are performed at nominal level  $\alpha = 5\%$.}
\label{fig:plot1}
\end{figure}

    Rejection frequencies are reported in 
Table~\ref{fig:plot1}. The left and center panels are  for the PBAR and NBAR models \eqref{BAR}, respectively, which are time-reversible. Rejection frequencies, thus,  should be less than the nominal size $5\%$. While the empirical size of the ICS method, irrespective of $n$, satisfies this level constraint, both  GHM strategies grossly violate it. The right panel is dealing with the QAR model \eqref{QAR}, the time-irreversibility of which is very well detected, for all $n$, by  the ICS method while the two GHM strategies yield extremely low powers. 

Whether they reject time-reversibility or not, the conclusions of the GHM tests, thus, should be taken with some care.


\section{Empirical study}\label{sec::ES}
We investigate the time-reversibility of climate change-related data previously analyzed in \cite{ghm22}. Table \ref{tab:1} provides a summary of the data, including abbreviations, observation periods, and observation frequencies (yearly or monthly). 
For GLO, GL, GO, SA, GHG, N2O, GCAG, GISTEMP, and GMSL, as in \cite{ghm22}, we apply the Hodrick--Prescott filter \citep{hp97}, with  smoothing parameter $\lambda=14400$ for monthly data and $\lambda=100$ for yearly data,  to remove trends. The remaining series (SOI, NAO, PDO, NH, SH) do not   exhibit any apparent trend. Table~\ref{tab:2} summarizes the sample sizes and the subsampling block sizes selected by \eqref{eqn:rt_bw}.
We apply the methods introduced in Section \ref{sec::TR} to   the detrended data (GLO, GL, GO, SA, GHG, N2O, GCAG, GISTEMP, GMSL) and the raw data (SOI, NAO, PDO, NH, SH).
The  $p$-values obtained  for the ICS method are reported in Table \ref{tab:3}. There is strong evidence of time-irreversibility for the GO (global ocean temperature anomaly) and SH (southern hemisphere sea ice area)  series.

In contrast, the GHM  method concludes that the  GO and SH series, but also the  GLO, GL, SA, and NAO ones are time-reversible, whereas time-irreversibility is detected in GHG, N2O, GCAG, GISTEMP, GMSL, SOI, PDO, and NH. No $p$-value is associated with these conclusions, though. The detail is below: 
GLO, GL, GO, SA, and SH are declared  time-reversible   because Gaussianity could not be rejected in Step~2, neither by  the Shapiro-Wilk test nor by the Jarque-Bera one. Non-rejection of  Gaussianity, however,   does not imply  Gaussianity, and these series  still could  be non-Gaussian and time-irreversible. In view of the ICS results, GO and SH may indeed fall into this case. 
For all other series,   Gaussianity  is rejected  in Step~2,  both by the Shapiro-Wilk and the Jarque-Bera test. For GHG, GCAG, GISTEMP, N2O, GMSL, and PDO, the causal and non-causal AR($r+s$) model orders selected by Student  maximum  likelihood in Step~3 do not satisfy $r=s$, hence the GHM method concludes to the time-irreversibility of these series. However, the possibility remains   that these processes are non-linear yet time-reversible.  
For SOI and NAO, the orders $r$ and $s$ selected by Student  maximum  likelihood in Step~3 are the same. While strategies 1 and 2 both conclude  that SOI is time-irreversible, there remains a possibility it is non-linear but time-reversible. 
Interestingly, BIC for the restricted likelihood  for NAO is smaller than BIC for the unrestricted one, and the equality of causal and non-causal parameters is not rejected by the likelihood ratio test in strategy 2, yielding the  conclusion  that NAO is time-reversible.  

The two methods, thus, yield quite different conclusions; while the reliability of the  ICS conclusions is controlled by a type one risk less than $5\%$, though, no risk evaluation  is possible for the GHM  method.

\begin{table}[h]
 \caption{\small List of datasets with their respective abbreviations, observation periods, and observation intervals}
 \label{tab:1}
 \centering
  \begin{tabular}{cccc}
data &abbreviation& period & observation interval\\\hline
 global land and ocean temperature anomaly&GLO&1881--2014&yearly\\
 global land temperature anomaly&GL&1881--2014&yearly\\
 global ocean temperature anomaly&GO&1881--2014&yearly\\
 solar activity&SA&1881--2014&yearly\\
 greenhouses gas&GHG&1881--2014&yearly\\
 nitrous oxide &N2O& 1881--2014&yearly\\
 global component of climate at a glance&GCAG&Jan. 1880--Dec. 2016&monthly\\
 global surface temperature change&GISTEMP&Jan. 1880--Dec. 2016&monthly\\
 global mean sea level&GMSL&Jan. 1880--Dec. 2013&monthly\\
 southern oscillation index&SOI&Jan. 1951--Dec. 2021&monthly\\
 north atlantic oscillation index&NAO&Jan. 1951--Dec. 2021&monthly\\
 pacific decadal oscillation index&PDO&Jan. 1854--Dec. 2021&monthly\\
 northern hemisphere sea ice area&NH&Jan. 1979--Dec. 2021&monthly\\
southern hemisphere sea ice area&SH&Jan. 1979--Dec. 2021&monthly\\\hline
  \end{tabular}
\end{table}

\begin{table}[h]
 \caption{Sample size for each index and the corresponding subsampling block size selected by \eqref{eqn:rt_bw}}
 \label{tab:2}
 \centering
  \begin{tabular}{ccc}
data  &sample size & subsamplig block size \\\hline
GLO&134&32\\
GL  &134&32\\
GO  &134&32\\
SA  &134&32 \\
GHG&134&32\\
N2O&134&32\\
GCAG&1644&256\\
GISTEMP&1644&256\\
GMSL  &1608&256\\
SOI    & 852  &128\\
NAO   & 852  &128\\
PDO   &  2016 &256\\
NH     & 516  &128\\
SH     & 516 &128\\
  \end{tabular}
\end{table}

\begin{table}[h]
\centering
\caption{\small $p$-values of the integrated copula spectrum (ICS)-based test by \cite{gkvvdh22} applied to climate-related data
}
 \label{tab:3}
\begin{tabular}{c|ccccccc}
Index & GLO & GL & GO & SA & GHG & N2O & GCAG \\\hline
$p$-value & 0.7480 & 0.9320 & 0.0291 & 0.2910 & 0.9900 & 0.9900 & 0.4470 \\[1ex]
Index & GISTEMP & GMSL & SOI & NAO & PDO & NH & SH \\\hline
$p$-value & 0.4270 & 0.7910 & 0.5630 & 0.7810 & 0.7790 & 0.1490 & 0.0000 \\
\end{tabular}
\end{table}

\newpage

\begin{acks}[Acknowledgments]
The authors are grateful to F.\ Giancaterini,  A.\  Hecq,  and C.\ Morana  for kindly sharing their R codes and climatic data. 
Yuichi Goto acknowledges support from  the JSPS Grant-in-Aid for Early-Career Scientists grant JP23K16851 and 
 thanks D.\ Paindaveine and T.\ Verdebout for supporting his  research stay at Universit\' e libre de Bruxelles where part of this research was conducted. Marc Hallin  gratefully acknowledges the support of the COST (European Cooperation in Science and Technology) Action HiTEc CA21163 and the Czech Science Foundation grants GA\v{C}R22036365  and GA24-100788. \end{acks}


\newpage

\bibliographystyle{apalike}
\bibliography{ref}

\end{document}